\documentclass{aa}  
\usepackage{graphicx}
\usepackage{natbib}
\bibpunct{(}{)}{,}{a}{}{;}
\usepackage{amssymb}
\usepackage{txfonts}
\begin{document}
   \title{Searching for Gas-Rich Disks around T Tauri Stars in Lupus}

 % \subtitle{CO observations with the JCMT}

   \author{T.A. van Kempen
          \inst{1}
          \and
          E.F. van Dishoeck\inst{1}
          \and
         C. Brinch\inst{1}
          \and
           M.R. Hogerheijde\inst{1}
}

   \offprints{T. A. van Kempen}

   \institute{$^1$Leiden Observatory, P.O. Box 9513,
              2300 RA Leiden, The Netherlands\\
              \email{kempen@strw.leidenuniv.nl}
                     }

   \date{}

% \abstract{}{}{}{}{} 
% 5 {} token are mandatory
 
  \abstract % context heading (optional) % leave it empty if necessary
   {}% aims heading (mandatory) 
 {To characterize the molecular
   environment of classical T Tauri stars in Lupus observed with the
   Spitzer Space Telescope and to search for gas-rich disks toward these
   sources.}
{Submillimeter observations of $^{12}$CO, $^{13}$CO and C$^{18}$O in
  the $J$=3-2 and 2-1 lines were obtained with the James Clerk
  Maxwell Telescope toward a sample of 21 T Tauri stars with disks in
  the Lupus molecular clouds. Pointings at the sources and at selected off-positions are presented in order to
  disentangle material associated with disks from ambient cloud
  material.}  
{One source, IM Lup (Sz 82), has been found with
double-peaked $^{12}$CO and $^{13}$CO profiles characteristic of a
  large rotating gas disk. The inclination of the disk is found to be $\sim 20^\circ$, with an outer radius of 400-700 AU. 
  For most other sources, including GQ Lup with its
  substellar companion, the single-dish $^{12}$CO lines are
  dominated by extended cloud emission with a complex 
  velocity structure.  No evidence for molecular outflows is found.
 Compact C$^{18}$O emission due to dense circumstellar
  material has been detected toward only two sources. Future searches
  for gas-rich disks in Lupus should either use interferometers or
  perform very deep single-dish integrations in dense gas tracers to
  separate the cloud and disk emission.}
   {}


   \maketitle
%
%________________________________________________________________

\section{Introduction}

A general characteristic of pre-main sequence (PMS) stars is the
presence of circumstellar disks, which have been observed for a wide
range of evolutionary stages associated with low-mass star formation
\citep[e.g.,][]{Lada87,Andre93,Greaves05}. These disks contain all the
necessary ingredients for forming complex planetary systems similar to
our own Solar system, now also seen around other stars
\citep[e.g.,][]{Ruden99,Marcy95}.  The gas-rich phase of disks is
crucial in the evolution, since gas is needed to build Jovian-type
planets. Also, the presence of gas affects the dynamics and evolution
of the dust in the system and the migration of any planets embedded in
the disk \citep[e.g.,][]{Gorti04}. With time, the disk will loose its
gas through processes such as photoevaporation, stellar winds,
formation of gas-rich planets and accretion onto the star
\citep[e.g.,][]{Hollenbach00,Alexander05}. However, the timescale for
gas dissipation in disks is still poorly constrained observationally
and it is even unclear whether gas and dust are lost simultaneously
from disks when they evolve from the massive optically thick to the
tenuous optically thin `debris' phase.

To constrain these timescales, surveys of gas in disks in a variety of
environments are needed with sensitivities sufficient to detect a Jupiter mass of gas ($\sim$10$^{-3}$ M$_\odot$).
%typically consisting of minimum masses present in the gas of a few
%times 10$^{-4}$ M$_\odot$. 
Surveys to date have focussed on CO millimeter observations, mostly of
disks in the Taurus molecular cloud \citep[e.g.][]{Thi01}.  Although
CO is known to be a poor tracer of the gas mass due to the combined
effects of photodissociation in the upper layers and freeze-out in the
midplane \citep[e.g.,][]{Zadelhoff03}, it is much simpler to observe
with current instrumentation than other tracers including H$_2$
\citep{Thi01} or atomic fine-structure lines \citep{Kamp03}.
Interferometer surveys have detected gas-rich disks around a large
fraction ($>$60\%) of classical T Tauri stars in Taurus-Auriga with
ages up to a few Myr and gas masses of at least a few times $10^{-4}$
M$_\odot$, depending on assumptions about the CO abundance
\citep{Koerner95,Dutrey96,Dutrey03}. In contrast, the detection rate
is only 10\% for the more evolved weak-line T Tauri stars
\citep{Duvert2000}. Similar statistics are found in a single-dish
submillimeter survey of CO in a spatially more distributed sample of
Herbig Ae and Vega excess stars \citep{Dent05} and in CO infrared
surveys \citep{Najita03,Blake04}. 

Surprisingly little is known about the presence of gas-rich disks around T Tauri stars in
other nearby star-forming regions such as Corona Australis, Lupus and
Chamaeleon. This is largely due to the absence of interferometers in
the Southern sky, and even Ophiuchus has been poorly sampled with
current facilities. We present here an initial search for gas-rich
disks in the Lupus molecular clouds, using a single dish telescope.

The Lupus clouds, located at around galactic coordinates 335$^\circ \le l
\le 341^\circ$ and $7^\circ \le b \le 17^\circ$ \citep{Hughes94}, are among the
closest star-forming regions to the Sun at $\sim$150 pc \citep[see
reviews by][]{Krautter91,Comeron06}.  Mid M-type stars dominate the
stellar population of Lupus, which contains no confirmed O or B-type
stars. In total 69 Pre-Main Sequence (PMS) objects were found by
\citet{Schwartz77}, distributed over four clouds now known as Lupus
1--4 \citep[see also][]{Hughes94}. In addition, some 130 new weak-line
T Tauri stars were detected in ROSAT images
\citep{Krautter94,Krautter97}. The most massive stars in the complex
are two A-type stars, HR 5999 and HR 6000, present in the richest
sub-grouping in stars and complexity, Lupus 3.
%Similar to the Taurus-Auriga clouds, the detection rate of
%stars in Lupus is 38 \%.  
A good overview of the clouds is given by the extinction map made by 
\citet{Cambresy99}, based on star counts. Recently a higher resolution ($\sim 30''$) map
has been published by \citet{Teixeira05}. 

 At millimeter wavelengths the Lupus clouds have been studied both in
molecular emission lines and in the continuum. $^{12}$CO
\citep{Tachihara01}, $^{13}$CO \citep{Tachihara96} and C$^{18}$O
\citep{Hara99} $J$=1--0 large-scale maps were made at 2\farcm6
resolution with the NANTEN telescope. Higher resolution observations
were done with the Swedish-ESO Submillimeter Telescope (SEST) over
more limited regions with $\sim$45$''$ resolution by \citet{Gahm93}
(mostly $^{12}$CO 1--0 in Lupus 2), \citet{Rizzo98} ($^{12}$CO 1--0 in
Lupus 1 and 4 filaments) and \citet{Vilas-Boas00} ($^{13}$CO and
C$^{18}$O 1--0 in localized dark cores in Lupus
1-4). \citet{Nuernberge97} performed a continuum survey of 32 T Tauri
stars in Lupus at 1.3 mm with the SEST, probing the cold circumstellar
dust. 
%Lupus 3 has recently recieved new attention in both gas and dust
%observations \citep{Tachihara05} and infrared extinction
%\citep{Teixeira05}.

The Lupus star-forming region appears to differ in several aspects
from other low-mass star-forming clouds. In contrast with Ophiuchus
and Taurus, which contain a significant fraction of deeply embedded
Class 0 and Class I objects, Lupus has at most a few embedded objects
\citep{Krautter91, Tachihara96, Comeron06}. This could be an
indication that Lupus is more evolved than other star-forming regions.
The peak in the age distribution is $3\times10^6$ yr for stars in
Lupus 3 and 4, calculated from the PMS evolutionary tracks of
\citet{dantona94} for a distance of 150 pc \citep{Hughes94}. Under the
same assumptions, Lupus 1 and 2 PMS stars are younger with a peak of
$1\times10^6$~yr. These age estimates are comparable to those for
other star-forming regions such as Chamaeleon \citep{Hartigan93}, but
are about double that of the Taurus cloud \citep{Simon93}.  Thus, a
study of disks in Lupus is interesting because this is precisely the
age range in which significant evolution of the gas disk is expected.

A second characteristic of the Lupus clouds is the distribution of
spectral types. Compared to the Taurus cloud, Lupus is dominated by
lower mass stars. The peak of the distribution in spectral type is M0,
with very few stars having spectral types higher than K7. In contrast,
the peak of the spectral type distribution in
Taurus is at K7. 

%dubbel op
%Mass distributions calculated from PMS evolutionary tracks by
%\citet{Dantona94} confirm that the Lupus clouds produces very few
%stars with solar masses and above compared to clouds such as Taurus
%and Chameleon. \citep{Hughes94} \\

In this work, single-dish submillimeter observations of CO toward a
sample of 21 classical T Tauri stars with disks are presented obtained
with a beam of 14$''$, much smaller than that of previous data.  The sources are
a subset of the PMS stars found by
\citet{Schwartz77}, \citet{Krautter91} and \citet{Hughes94}, and they have been observed
with the {\it Spitzer Space Telescope} in the context of the `From Molecular Cores to Planet Forming Disks' (c2d) Legacy survey \citep{Eva03}.  The aim of the CO
observations is to characterize the molecular environment of the T
Tauri stars at high angular resolution and to search for gas-rich
disks suitable for future follow-up interferometer observations.

%___________________________________________________________________

\section{Observations}

\subsection{CO observations}

The submillimeter CO observations were carried out in two runs in
April and July 2005 at the James Clerk Maxwell Telescope
(JCMT)\footnote{The JCMT is operated at the summit of Mauna Kea,
Hawaii by the Joint Astronomy Center on behalf of the United Kingdom
Particle Physics and Astronomy Research Council, the Netherlands
Organization for Scientific Research, and the National Research
Council of Canada}.  $^{12}$CO $J$=3--2 and C$^{18}$O $J$=3--2 and 2--1
transitions (depending on weather conditions) were observed with the
dual polarization B3 and single polarization A3 receivers.  The beam
sizes are 14$''$ and 20$''$, respectively. The Digital Autocorrelator
Spectrometer was used as the back-end with a bandwidth of 125 MHz,
giving a spectral resolution of 98 kHz for the A-receiver
corresponding to 0.10 km s$^{-1}$ at 230 GHz, and 196 kHz for the
dual-polarization B-receiver corresponding to 0.22 km s$^{-1}$ at 345
GHz.  The integration times were such that the RMS
%for $^{12}$CO 3--2 lines at a level of 0.2
%K observable with S/N of around 10-15, 
is around 70--100 mK in a 0.2 km s$^{-1}$ velocity bin for the
B-receiver and 50 mK for the A-receiver, depending on source and observing
conditions, on a T$_{\rm{mb}}$ scale. Pointing was checked regularly for these southern sources
and found to vary within 2$''$. Beam efficiencies were taken to be
0.65 for RxA3 and 0.63 for RxB3\footnote{Observed by the JCMT staff
over many observations; see the JCMT website
http://www.jach.hawaii.edu/JCMT/}.

Frequency switching with a switch of 16.4 MHz was adopted for $^{12}$CO because no emission-free
position could readily be found for many of the sources in our
sample. If a signal was detected on source, additional observations
were carried out at 30$''$ offsets to the east and south of the source
to characterize its environment and constrain the on-source
contribution.  If indications were found of a gas-rich disk,
additional follow-up observations were done on-source in $^{13}$CO
3--2 using a beam switch of 180$''$ in AzEl.  C$^{18}$O 3--2 or 2--1
observations were performed for nearly all sources, also using a beam
switch of 180$''$.  Table~1 summarizes the observed lines per source.
The data were reduced with the SPECX and CLASS reduction packages.

Our observing strategy is similar to that for the single-dish
observations of gas-rich disks in Taurus by \citet{Thi01}. Because the
surface layers of disks are warm, the $^{12}$CO 3--2 line is well
suited to observe these regions. Also, the smaller beam size at higher
frequencies is more favorable for observing disks, which have typical
sizes of a few hundred AU corresponding to a few arcsec at the
distance of Lupus. Typical $^{12}$CO 3--2 antenna temperatures of
disks in Taurus are 0.5--1 K, so that an RMS of $<$0.1 K needs to be
reached.  To distinguish gas associated with disks from that of
surrounding clouds, observations on-source and at nearby off-source
positions need to be taken.  Gas-rich disks should show a
characteristic double-peaked line profile and a stronger signal
on-source than at the off-source positions, while cloud emission
should produce a comparable signal at all positions.  Finally, disks
in Taurus have not been detected in C$^{18}$O with single-dish
telescopes. On the other hand, such beam-switched C$^{18}$O
observations have been found to be a good probe of any dense
circumstellar gas or remnant envelope associated with the source. For
example, Class I sources show typical C$^{18}$O intensities of 0.5--2
K \citep{Hogerheijde98,Jorgensen02}.

%
%________________________________________________________________

\subsection{Sample}

Our sample of classical T Tauri stars in Lupus was selected from
targets belonging to the InfraRed Spectrometer (IRS) sample of the c2d
Legacy program on {\it Spitzer} (See Table 1).  They are a subset of
the T Tauri stars studied by \citet{Schwartz77}, \citet{Krautter91} and
\citet{Hughes94}, selected to be brighter than 200 mJy at 12 $\mu$m
and excluding sources contained in the {\it Spitzer-IRS} guaranteed
time programs.  Infrared photometry and spectroscopy was performed on
all sources giving estimates of their spectral types (see Table 1).
In total 21 PMS stars were observed, most of them located in Lupus 3.
All sources show a clear 10 $\mu$m emission feature due to warm
silicates, confirming the presence of a circumstellar disk
\citep{Kessler06}.  Some of the selected stars are known
binaries. Where available, results on the 1.3 mm continuum emission are
reported \citep{Nuernberge97}. This emission is presumably dominated
by the cold dust in the disk.

The distances to individual Lupus clouds have considerable
uncertainty.  \citet{Hughes93} found 140$\pm$20 pc from spectroscopic
parallax of field stars around the clouds.  However, observations of
reddening and interstellar absorption lines toward individual stars by
e.g., \citet{Knude98} and \citet{Crawford00} find distances ranging
from 100 to 200 pc.  A good discussion can be found in
\citet{Comeron06}. The depth of the Lupus clouds may be non-neglible
and can be as large as 50 pc. Also, it has been suggested that not all clouds of
Lupus are related, especially Lupus 2 could be located as far as 360
pc \citep{Knude01}. However this seems unlikely given the narrow
ranges in mean velocity of the clouds (see e.g., Tachihara et al.\
1996, Vilas-Boas et al.\ 2000 and this work $\S 3$).  The distance
used here for all sources is that most commonly adopted, 150 pc.

%________________________________________________________________

\begin{figure*}[!pht]
   \centering
   \includegraphics[width=400pt]{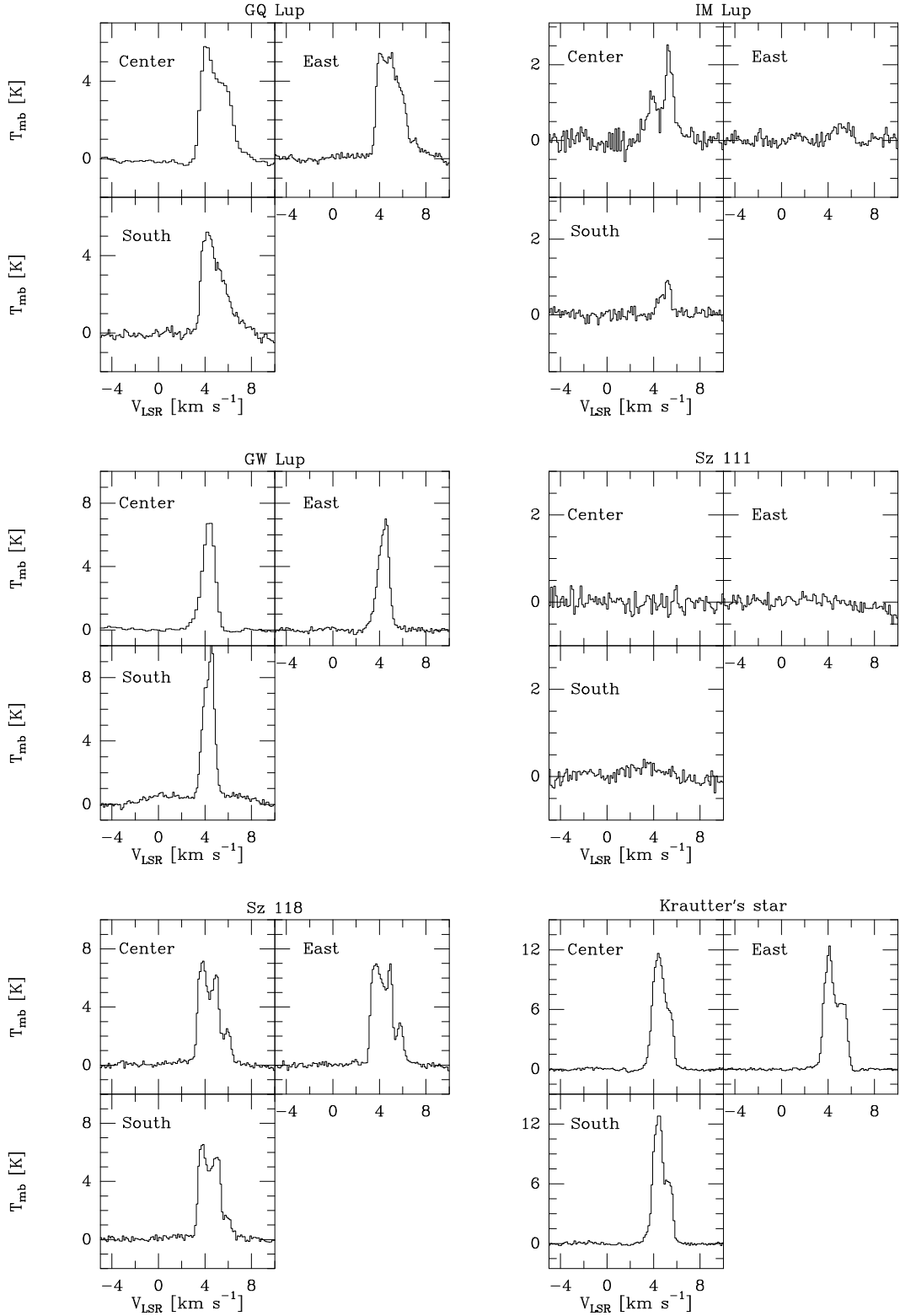}
      \caption{$^{12}$CO  3--2 lines of selected sources in Lupus, 
with off-positions 30$''$ south and east of source. }
         \label{Fig:CO}
   \end{figure*}

\begin{table*}[!htp]
%\begin{minipage}[t]{\columnwidth}
\label{sample}      
%\centering        
 \caption{Summary of the properties of T Tauri stars and observed lines in Lupus}\renewcommand{\footnoterule}{}                     
\begin{tabular}{ c c c c c c c c c c}     % 7 columns 
\hline       \hline
Source   & RA  (J2000)  & Dec (J2000)    & Spectr.type$^a$ & Cloud & $F_{1.3\rm{mm}}$(mJy)$^b$ & $^{12}$CO & Offsets &C$^{18}$O & Add. obs \\ \hline 
\object{HT Lup$^c$ } & 15 45 12.87 & $-$34 17 30.6 & K2 & Lupus 1 &135 & 3--2 & E S &3--2/2--1& \\ 
\object{GW Lup  }  & 15 46 44.68 & $-$34 30 35.4 & M2-M4 & Lupus 1 &106 &3--2 & E S&3--2/2--1& \\ 
\object{HM Lup  }  & 15 47 50.63 &$-$35 28 35.4 & M3 & Lupus 1&$<$45 &3--2& E S &3--2/2--1& \\ 
\object{Sz 73}     & 15 47 56.98 & $-$35 14 35.1 &M0 & Lupus 1& 26 & 3--2 & E S&2--1& \\ 
\object{GQ Lup   } & 15 49 12.12  &$-$35 39 05.0 & K7-M0 & Lupus 1& 38 & 3--2 & E S &2--1 & $^{13}$CO 3--2\\ 
\object{IM Lup   } & 15 56 09.17 & $-$37 56 06.4 & M0 & Lupus 2&260 & 3--2 & E S& 2-1& $^{13}$CO 3--2\\ 
\object{RU Lup}    & 15 56 42.31 &$-$ 37 49 15.5 & K7-M0 & Lupus 2& 197 &3--2 & E S&3--2 & \\ 
\object{Sz 84}     & 15 58 02.52 & $-$37 36 02.8 &M5.5 & Lupus 2& $<$36 & 3--2 & E S&2--1& \\ 
\object{RY Lup}    & 15 59 28.39 & $-$40 21 51.2 & K4 & Lupus 3& -& 3--2 & - & 3--2& \\ 
\object{EX Lup   } & 16 03 05.52 & $-$40 18 24.9 & M0 & Lupus 3  &-  & 3--2 & - &3--2 &  \\ 
\object{HO Lup$^c$}  & 16 07 00.61 &$-$39 02 19.4 & M1 & Lupus 3&$<$24 &3--2 & E S &3--2 & \\ 
\object{Sz 96}     & 16 08 12.64 & $-$39 08 33.3 &M1.5 & Lupus 3& $<$45 & 3--2 & E S& 3--2& \\ 
\object{Krautter's star}&16 08 29.70 &$-$39 03 11.2 & K0 & Lupus 3& $<$30 & 3--2 & E S&3--2 & \\ 
\object{Sz 107}    & 16 08 41.78 & $-$39 01 36.4 & M5.5 & Lupus 3&-  & 3--2 & E S& - & \\ 
\object{Sz 109}    & 16 08 48.17 & $-$39 04 19.0 & M5.5 & Lupus 3&- & 3--2 & E S& 3--2&\\ 
\object{Sz 110}    & 16 08 51.56 & $-$39 03 17.5 & M2 & Lupus 3& - &  3--2 & E S & 3--2&\\ 
\object{Sz 111}    & 16 08 54.74 & $-$39 37 17.5 & M1.5 & Lupus 3& -& - & - & 3--2&\\ 
\object{V908 Sco} & 16 09 01.84 & $-$39 05 12.1 & M4 & Lupus 3& -&  3--2 & E S &3--2 &\\ 
\object{Sz 117}    & 16 09 44.34 & $-$39 13 30.4 & M2 & Lupus 3& -& 3--2 & E S & 3--2&\\ 
\object{Sz 118}    & 16 09 48.68 & $-$39 11 17.2 & K6 & Lupus 3& -& 3--2 & E S& 3--2&\\ 
\object{Sz 123}    & 16 10 51.58 & $-$38 53 13.7 & M1 &Lupus 3 & -& 3--2 & E S& 3--2&\\ 
\hline 
\end{tabular}\\
$^a$ \citet{Hughes94} \\
$^b$ \citet{Nuernberge97}, where available\\
$^c$ Known binaries \citep{Nuernberge97} \\
%\end{minipage}
\end{table*}

%___________________________________________________________________

\section{Results}
\subsection{$^{12}$CO}

The observed $^{12}$CO  3--2 line intensities are presented in
Table~2, together with the peak intensities. Typical examples of spectra, illustrating the range of observed profiles, both on-source as well as at off-source positions, are shown
in Fig. \ref{Fig:CO}. The remainder of the spectra can be found in the
on-line material.

Only a single source is found in our sample with a clear signal of gas
associated with the source itself, namely \object{IM Lup}. A classic
double-peaked disk profile is found toward the source position, which
is largely absent at the 30$''$ east and south positions offset from
the source (see Fig. \ref{Fig:CO}). The profile, with a maximum peak
of 2.5 K, is significantly broader than that associated with
quiescent gas. This source is further discussed in \S 4.1.  

None of the other sources show indications of gas-rich disks surrounding the central
T Tauri star. Three sources, RY Lup, EX Lup and Sz 111, appear to be located outside any
cloud material and have no detectable disk emission at our sensitivity
limit. Using the formula in \citet{Thi01} with the observed limits on $^{12}$CO emission, upper limits of the gas masses of $1\times10^{-6}$ M$_\odot$ are obtained for these disks with a CO/H$_2$ ratio of $10^{-4}$. For a more realistic overall CO/H$_2$ disk abundance of 10$^{-6}$, the limits would increase to $1\times10^{-4}$ M$_\odot$ (see also \S 4).

 Most sources have a strong $^{12}$CO line on source, but
these signals are also detected at the off-positions with comparable
strengths (see Fig. \ref{Fig:CO}). These lines must originate in the
large scale molecular cloud material in the line of sight of the
source. The complex profiles seen for many sources indicate a complex
structure of different molecular clouds moving at different
velocities, either in the fore- or background.  For example, the
spectra toward Sz 118 (see Fig. \ref{Fig:CO}) show three peaks with
different intensities at all positions with widths of $\sim$0.8 km
s$^{-1}$ and velocities that differ by 1--1.5 km s$^{-1}$. C$^{18}$O
is not detected at all for this source.  Comparison with the large
scale maps shows that observations of $^{12}$CO 3--2 with peak
temperatures of 2--20 Kelvin are consistent with the lower resolution
$^{12}$CO surveys \citep{Gahm93, Tachihara96}. 
%The line intensities correspond typical H$_2$ column densities of a few
%times 10$^{21}$ cm$^{-2}$ corresponding to $A_V\approx$ a few mag,
%consistent with extinction maps by \citet{Cambresy99}.

The $^{12}$CO line widths of $\sim$0.8 km s$^{-1}$ are comparable to
those found in \citet{Vilas-Boas00} and \citet{Tachihara01}, taking
into account multiple clouds at different velocities.  No evidence for
line wings at $\geq$2 km s$^{-1}$ from line center is found at the
level $T_{\rm{MB}}<0.15$~K (2$\sigma$), providing limits on any
small-scale molecular outflows associated with these T Tauri
stars. For RU Lup, this is consistent with \citet{Gahm93}, who
reported no outflows from CO 1--0 data in a larger beam.

Although no direct correlation can be inferred due to the confusion of cloud emission around most sources, it is interesting to note that the largest 1.3 mm flux found by \citet{Nuernberge97} is associated with IM Lup. Thus it also contains the most surrounding dust, in addition to being the only source with CO disk emission.

%
%________________________________________________________________

%
%________________________________________________________________

\subsection{C$^{18}$O and $^{13}$CO}
The results for the C$^{18}$O and $^{13}$CO line intensities are
included in Table~2. $^{13}$CO 3--2 was observed for only two objects: IM
Lup and GQ Lup. A more detailed description for these two objects can
be found in \S 4. Figure 2 shows some examples of observed
C$^{18}$O lines. 
 
%They can be assumed to be mostly optically thin in
%interstellar or circumstellar conditions such as found around young
%protostars and in molecular clouds \citep[e.g.,][]{Jorgensen02}. A
%large signal of C$^{18}$O such as detected toward HT Lup points to a
%compact cloud with a large column density, possibly a dense
%circumstellar envelope.

Only two sources in our sample show a C$^{18}$O line above the noise
level: \object{HT Lup} and \object{Sz 73}.  HT Lup is the only source
toward which both C$^{18}$O 2--1 and 3--2 lines have been detected in
our survey.  Because the data were taken in 180$''$ beam-switched
mode, the bulk of the emission is likely  directly associated with the source. This could point towards a small circumstellar envelope.  The
2--1/3--2 line ratio of $>1$ indicates either very cold gas ($<15$~K)
or densities less than $10^5$ cm$^{-3}$ for $T\approx 20$~K. The
inferred C$^{18}$O column density for these conditions is $\sim
10^{15}$ cm$^{-2}$, corresponding to a CO column density of $5\times
10^{17}$ cm$^{-2}$ for $^{16}$O/$^{18}$O=540 \citep{Wilson94} and an
H$_2$ column density of $6\times 10^{21}$ cm$^{-2}$ for a dark cloud
C$^{18}$O abundance of $1.7\times 10^{-7}$ \citep{Frerking82}. Thus,
the extinction directly associated with the source is $A_V\approx 6$
mag using $N_{\rm H}/A_V=1.8\times 10^{21}$ cm$^{-2}$ mag$^{-1}$
\citep{Rachford02}.

As mentioned before, our observations are not sensitive to extended
C$^{18}$O emission because of beam-switching by 3$'$.  The lack of any
features at the $2\sigma$ level of $<$0.1 K in C$^{18}$O 2--1 and
$<$0.3 K in 3--2 implies that either any large scale cloud emission
must be very smooth on 3$'$ scale or that there is no C$^{18}$O at
this limit.  \citet{Vilas-Boas00} and \citet{Hara99} find C$^{18}$O
1--0 antenna temperatures of 0.2 to 1.5 K for the densest cores.  For
typical dark cloud conditions, the 2--1 line should be comparable in
strength whereas the 3--2 line should be factors of 2--4 weaker, so
that these lines would have been detectable at the positions of
strongest 1--0 emission. However, most of our sources are located at
the edge or outside the C$^{18}$O contours of \citet{Hara99}
suggesting that C$^{18}$O is indeed largely absent.  The lack of
C$^{18}$O 2--1 emission at the level of 0.08 K ($2\sigma$) indicates a
C$^{18}$O column density of less than $5\times 10^{13}$ cm$^{-2}$, or
H$_2$ less than $3\times 10^{20}$ cm$^{-2}$ using the dark cloud
abundance. In this diffuse cloud regime, however, the C$^{18}$O
abundance may be significantly lower than the standard dark cloud
value \citep{VanDishoeck88}, increasing the H$_2$ column density.

%
%________________________________________________________________

  \begin{figure}[!htp]
  \centering
       \includegraphics[angle=270,width=250pt]{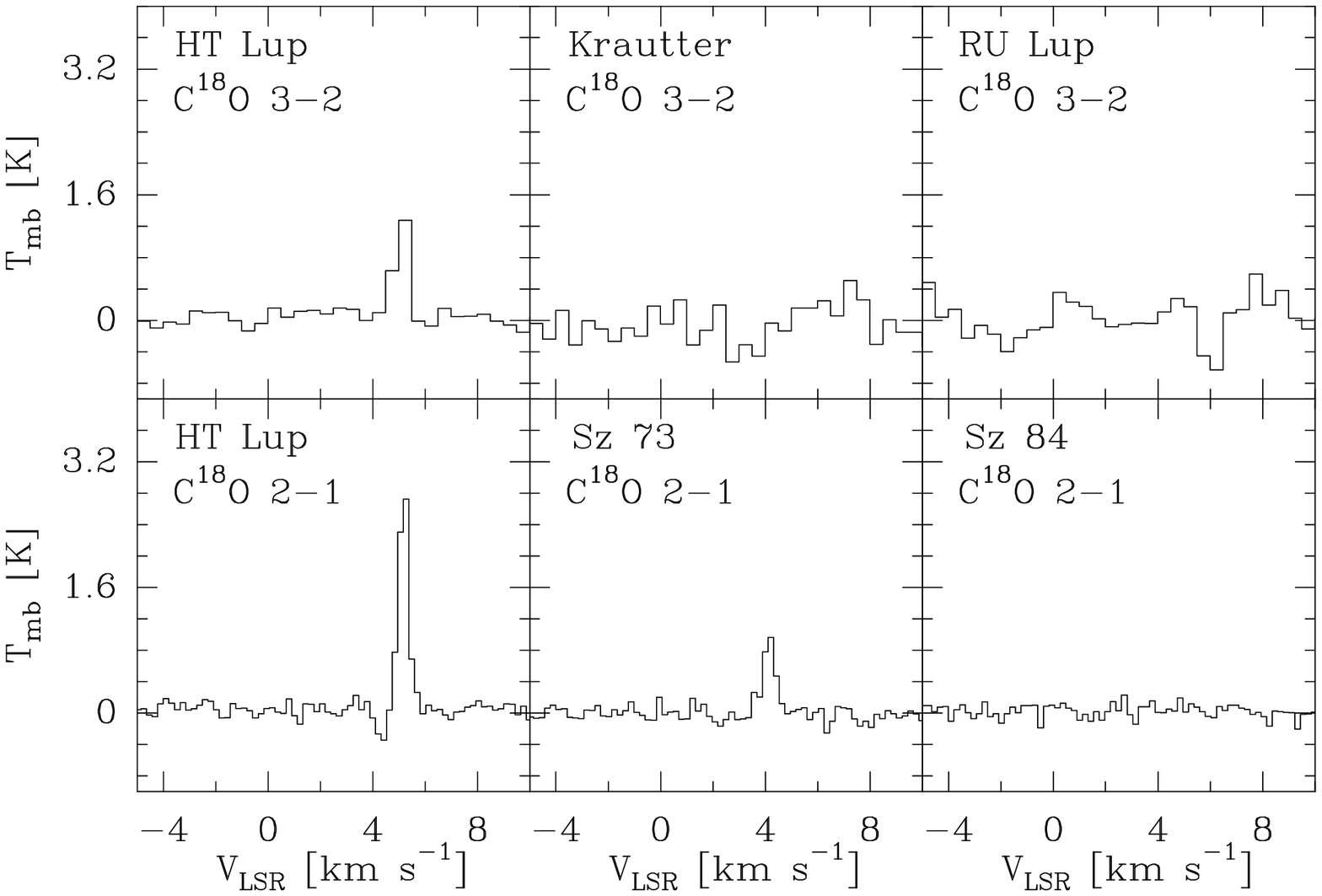}
      \caption{C$^{18}$O lines toward selected sources in Lupus. }
              \label{FigC18O1}
   \end{figure}
%
%________________________________________________________________

%
%________________________________________________________________

\begin{table*}
%\begin{minipage}[t]{\columnwidth}
\caption{Observed line intensities toward T Tauri stars in Lupus} 
\label{table:results}      
%\centering   
   
% \renewcommand{\footnoterule}{}              
\begin{tabular}{c c c c c c c c}     % 7 columns 
\hline \hline       
Source   & \multicolumn{2}{c}{$\int T_{\rm{MB}}dV$}  &  &  \multicolumn{4}{c}{Peak $T_{\rm{MB}}$} \\\cline{2-3} \cline{5-8}
 &    $^{12}$CO 3--2    &  $^{13}$CO 3--2      &    & $^{12}$CO         &C$^{18}$O 3--2$^a$     &  C$^{18}$O 2-1$^a$& $^{13}$CO 3--2\\ 
         & [K km s$^{-1}$] & [K km s$^{-1}$] & &[K]               & [K]                  & [K]               & [K] \\ \hline 
\object{HT Lup }   & 26.5  &    &  & 15  & 1.9 & 3.5 &\\
\object{GW Lup}    & 9.1    &   &  & 6.7 & $<$ 0.16&$<$ 0.08 &\\
\object{HM Lup }   & 6.4   &    &  & 5.2 &$<$ 0.15 &$<$ 0.08 &\\
\object{Sz 73    } & 15.8    & &   & 9.0 & & 1.0 &\\
\object{GQ Lup   } & 13.4 & $<$0.14$^c$ &  & 5.8 & &$<$0.05 & $<$0.25$^c$\\ 
\object{IM Lup   } & 4.7  & 0.52    &   & 2.4 & &$<$0.09 & 0.35\\
\object{RU Lup   } &     -   &  &   & -  & $<$0.27& &\\
\object{Sz 84    } & 0.6     & &   & 0.8& & $<$0.09&\\
\object{RY Lup   } &  $^b$  & &  & $<$0.25   &$<$0.31 & &\\
\object{EX Lup   } & $^b$ &  & & $<$ 0.16 &$<$0.40& &\\
\object{HO Lup }   & 1.3   &    &  & 0.8 &$<$ 0.17 & &\\
\object{Sz 96    } & 25.1    & &   & 12 & $<$ 0.16& &\\
\object{Krautter's star}& 17.5 & & & 11.7 &$<$0.25 & &\\
\object{Sz 107  }  & 18.3    & &   & 12 & & &\\
\object{Sz 109  }  & 34.8    & &   & 26 &$<$ 0.15  & &\\
\object{Sz 110  }  & 26.1   & &   & 21 &$<$ 0.25 & & \\
\object{Sz 111  }  &  $^b$     &  &  & $<$ 0.2  &$<$0.30 & & \\
\object{V908 Sco} & 22.6    & &   & 13.7  &$ <$ 0.26 & & \\ 
\object{Sz 117   } & 11.4   &  &   & 5.5 &$<$ 0.22 &  &\\
\object{Sz 118   } & 14.3   &  &   & 7.2 &$<$ 0.27 & & \\
\object{Sz 123   } & 2.4    &  &   & 1.6 &$<$0.20 & &\\
%\object{IRAS 17159-4324}& 18.0 & & 13.0  &$<$ 0.22 & &\\
%\object{IRAS 17193-4319}&44.3&  & 12.1  & 2.4 & &\\
\hline
\end{tabular} \\
$^a$ If no line was detected, a 2$\sigma_{rms}$(see $\S$ 2.1) limit
in $\delta v = 0.25$ km s$^{-1}$ is given.\\
$^b$ No detection. See peak value for RMS\\
$^c$ see $\S$ 4.2\\
%\end{minipage}
\end{table*}

%________________________________________________________________

\section{Individual objects}
   \begin{figure}
  \centering
       \includegraphics[width=230pt]{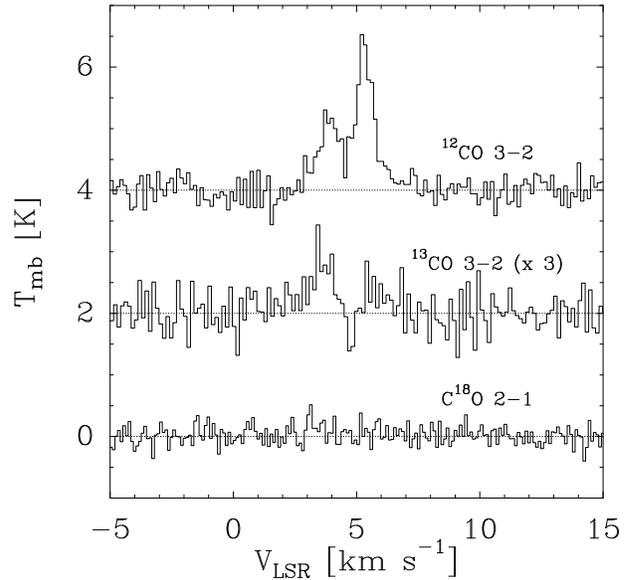}
      \caption{$^{12}$CO 3--2, $^{13}$CO 3--2 and C$^{18}$O 2--1
spectra observed toward IM Lup (Sz 82) shifted by +4, +2 and 0 K,
respectively.}
              \label{Figimlup}
   \end{figure}
\subsection{\object{IM Lup}}

\object{IM Lup} is the only T Tauri star in our sample with a clear
detection of $^{12}$CO emission associated with the source.  A
double-peaked profile with peak separations of about 2 km s$^{-1}$ is
seen in both $^{12}$CO and $^{13}$CO, consistent with a rotating gas
disk (see Fig.\ 3).  In addition, the lack of C$^{18}$O emission over
0.09 K in a 0.25 km s$^{-1}$ bin implies the absence of a significant contribution of a dense
circumstellar envelope.  The existence of a large disk around IM Lup
has recently been shown in scattered light images at visible and
near-infrared wavelengths with the {\it Hubble Space
Telescope} (Padgett et al, priv. comm. Schneider et al, priv. comm.).

The $^{12}$CO peak intensity for IM Lup is comparable to that of the
brighter sources with gas-rich disks in Taurus observed by
\citet{Thi01}. The much weaker ambient cloud material seen at the south off-position
(see Fig. 1) modestly affects the disk profile, however, especially the red
peak at $V_{\rm{LSR}}=$ 5 km s$^{-1}$. The blue peak at $V_{\rm{LSR}}=$ 3 km s$^{-1}$ is not
contaminated and is taken as the actual disk signal for both
peaks. The $^{13}$CO peaks of 0.2 K are comparable to those for Taurus
disks and do not seem to be affected by ambient cloud material

The observed $^{12}$CO/$^{13}$CO integrated intensity ratio of 6.0
$\pm$ 1.5 indicates $^{12}$CO and $^{13}$CO optical depths of 2.3 and
0.15, respectively, assuming that the excitation temperatures are the
same for both isotopes and constant throughout the disk.  The
uncertainties are dominated by calibration errors. If the peak
temperatures of the blue wing are used (1.5 for $^{12}$CO and 0.4 K
for $^{13}$CO respectively), optical depths of 3 and 0.3 are found.

\citet{Thi01} give several methods to derive disk masses from
different observations.  Assuming that most of the $^{13}$CO 3--2 emission
is optically thin and originates in the cold gas of the outer disk
with a single excitation temperature, the total gas mass is given by

\begin{equation}
M_{\rm{gas}} = 3 \times 10^{-6} \bigg( \frac{[^{12}\rm{CO}]/[^{13}\rm{CO}]}{60} \bigg) \bigg( \frac{\rm{H}_2/^{12}\rm{CO}}{10^4} \bigg) 
\end{equation}
\begin{displaymath}
\qquad \bigg( \frac{T_{\rm{ex}}+0.89}{e^{-16.02/T_{ex}}} \bigg) \bigg( \frac{\tau}{1-e^{-\tau}}\bigg) \bigg( \frac{d}{100 pc} \bigg) ^2 \int T_{mb} dV \quad M_{\odot}
\end{displaymath}

The H$_2$/CO and $^{12}$CO/$^{13}$CO ratios have been set to 10$^4$
and 60, following standard conversion factors.  This results in a
derived gas mass of $2\times 10^{-4}$ M$_{\odot}$ for IM Lup at
$d=150$ pc, using $T_{\rm{ex}} \simeq 30$ K.  For comparison, the disk mass found from the 1.3 mm
continuum flux \citep{Nuernberge97} is $6\times 10^{-2}$ M$_{\odot}$
assuming a gas/dust ratio of 100, $T_{\rm{dust}} =30$ K and a dust
opacity coefficient $\kappa$ of $0.01$ $\rm{cm}^{2}\rm{g}^{-1}$. The
factor of 300 between these two numbers is comparable to that found
for other disks \citep[e.g.,][]{Dutrey96,Mannings00,Thi01} and is
commonly explained by the fact that CO is not a good tracer of the gas
mass due to the combined effects of photodissociation by stellar and
interstellar UV radiation and freeze-out of CO onto the grains in cold
regions of the disk  \citep{Zadelhoff01,Aikawa02}.

The $^{12}$CO profile can be used to obtain estimates of two other
important disk parameters, the disk size and inclination.  Assuming 
optically thick $^{12}$CO 3--2 emission and following the method described
in \citet{Dutrey97}:

\begin{equation}\label{mass}
\int T_{\rm{mb}} \textrm{d} v \quad=\quad T_{\rm{ex}} (\rho \delta v) \quad \bigg( \frac{\pi(R^2_{out}-R^2_{in})}{d^2}\cos i \bigg)\quad \Omega^{-1}_a
\end{equation}
where $R_{\rm{in}}$ and $R_{\rm{out}}$ are the inner and outer radii,
$\delta v$ the turbulent velocity and $\rho$ a geometrical
factor. $\Omega^{-1}_a$ is the telescope beam on the sky and $i$ the
inclination, with 0 degrees being face-on with respect to the line of
sight. For IM Lup, $R_{\rm{in}}= 0 $ AU, a mean disk excitation
temperature $T_{\rm{ex}}$ of 25, 40 and 50 K and $ (\rho \delta v)$
of 0.3 km s$^{-1}$ are chosen.

The disk sizes calculated from Eq. (2) are given in
$R_{\rm{out}}\sqrt \cos i$. Scattered light images indicate that the inclination is of order 10$^\circ$ to 30$^\circ$, so that $\cos
i \approx 1$ is a reasonable approximation.  The disk sizes found with
$T_{\rm{ex}}=25$ and $T_{\rm{ex}}=40$ K are 730 and 580 AU,
respectively. Such numbers are comparable to the observed radius of
the dust disk in scattered light, about 4.3$''$ or 600 AU
(Padgett et al., priv. comm., Schneider et al.,priv. comm.)  \\

To better understand the observed line profile of $^{12}$CO, a simple
disk model has been used together with a ray-tracing program
\citep{Hogerheijde00} to compute the line profile.  Such models are
better suited to constrain parameters using a $\chi^2$
method with the observed spectrum. 
%with the observed line corrected
%for the observed ambient material. 
The populations of the $^{12}$CO rotational levels are assumed to be
in LTE throughout the entire disk.  The surface density of the disk is
defined as $\Sigma(r)=\Sigma_0(r/1$ AU)$^{-1.5}$  with a typical value
of $4.57 \times 10^{26}$ cm$^{-2}$ at 1 AU. The disk is assumed to be isothermal in the
vertical direction. The abundance of CO is set to
be 5$\times$10$^{-5}$ with respect to H$_2$, except in areas with temperatures
below 30 K where CO is frozen out onto the grains. There, abundances are taken to be 5$\times10^{-10}$. 
%These abundances are derived using the physical non-isothermal distribution in the vertical direction of the disk, which could not be included into the model. 
It is assumed
that the gas is in Keplerian orbits around a 0.5 $M_{\odot}$ star.  The
inner radius is set at 0.1 AU. The radial temperature profile is only constrained by a radial exponent of $-0.5$ and the luminosity.  The turbulent velocity is taken to be 0.1 km
s$^{-1}$, following \citet{Thi01,Qi04}.

The parameters to be fitted are the inclination, the outer radius of the disk and the temperature at the
inner radius.  The best fit is presented in Fig 4. It has a best fit
outer radius between 400 and 700 AU, consistent with the result of 600
AU found above.  The line profile is not very sensitive to the outer
radius or the CO abundance in the outer areas because temperatures are
too low to have a significant impact on the 3--2 emission.  The
observed line is quite sensitive to the inclination of the disk, which
is constrained to 20$^\circ$ $\pm$ 5$^\circ$. The temperature at the inner radius is taken to be 1200 K, but model results are not sensitive to the precise choice of temperature. 
%\citet{Padgett05} and \citet{Schneider05} propose an inclination of 20$^\circ$
\begin{figure}
   \centering
      \includegraphics[angle=270,width=230pt]{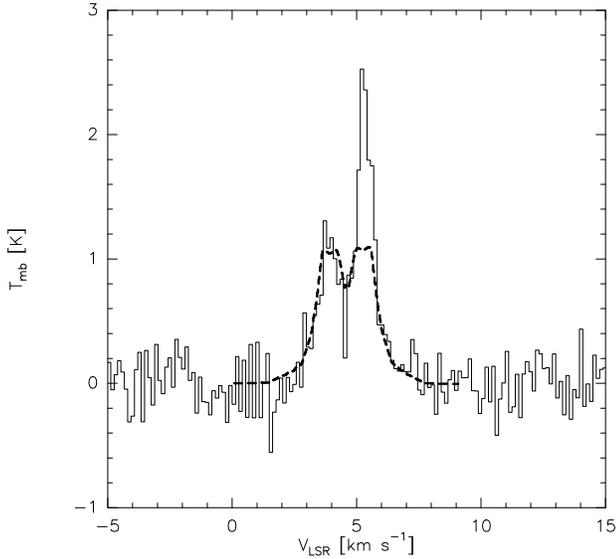}
      \caption{Comparison of the observed $^{12}$CO $J$=3--2 line toward IM Lup
with the best model fit with a disk of $i$=20$^\circ$, $R$=600 AU.}
%(****more work?)}
         \label{fits}
   \end{figure}
%________________________________________________________________

\subsection{GQ Lup}
   \begin{figure}
  \centering
       \includegraphics[width=230pt]{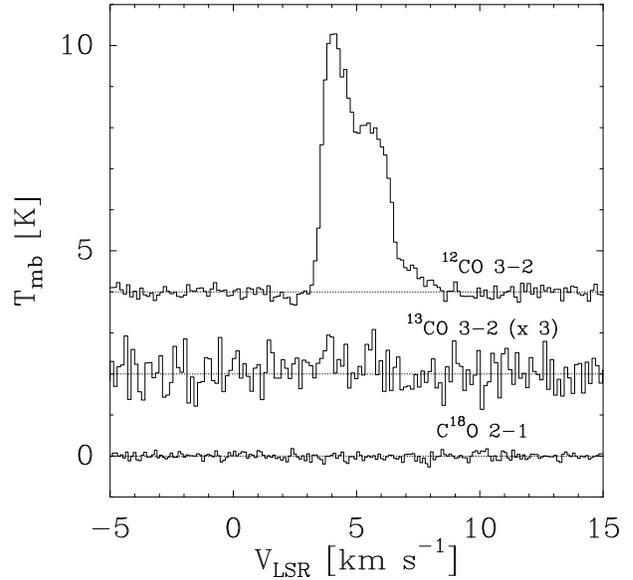}
      \caption{$^{12}$CO  3--2, $^{13}$CO 3--2 and C$^{18}$O 2--1 lines toward 
GQ Lup, shifted by +4, +2 and 0 K respectively.}
              \label{Fig:gqlup}
   \end{figure}

The classical T Tauri star GQ Lup has recently attracted attention
because adaptive optics imaging has revealed a substellar companion,
perhaps a massive Jupiter-like planet, with a separation of $\sim$100
AU \citep{Neuhauser05}. As for all objects in our sample, the GQ Lup
spectrum shows an infrared excess and a silicate emission feature at
10 $\mu$m, indicating the presence of a dust disk.  This object is
potentially interesting to further our understanding of the influence
of a companion on the disk structure and evolution.  High signal to noise
observations were taken toward this object (Fig.~\ref{Fig:gqlup}),
but like most other sources in our sample only diffuse interstellar
cloud material is detected in $^{12}$CO, as evidenced in the
observations at off-positions.  The $^{13}$CO spectrum shows a hint of
two peaks centered around $V_{\rm LSR}$=5 km s$^{-1}$ of equal height
at $T_{\rm mb}$=0.25 K, but deeper integrations are needed to confirm
this. C$^{18}$O 2--1 was not detected at all down to 5 mK
(2$\sigma$) in a 0.25 km s$^{-1}$ bin.  This constrains the maximum column density of C$^{18}$O
directly associated with the source to $\sim 2\times10^{12}$ cm$^{-2}$. An unresolved disk would then be limited to a gas mass of $<2.3\times10^{-5}$ M$_\odot$ for C$^{18}$O/H$_2=1.67\times10^{-7}$ or $<2.3\times10^{-3}$ M$_\odot$ for a lower ratio of C$^{18}$O/H$_2=1.67\times10^{-9}$. The mass derived from the 1.3 mm flux of 38 mJy is 2.1$\times10^{-3}$, assuming a gas/dust ratio of 100, $T_{\rm{dust}}=30$ K and a dust opacity $\kappa$ of 0.01 cm$^2$g$^{-1}$\citep{Nuernberge97}.

%___________________________________________________________________

\section{Concluding remarks}

The main conclusions of this work are as follows:
\begin{description}
\item{--} A $^{12}$CO $J$=3--2 survey of 21 classical T Tauri stars
with dust disks in Lupus taken in frequency-switched mode has revealed
mostly extended diffuse cloud material with a few magnitudes of extinction
and complex velocity structure.

\item{--} A C$^{18}$O $J$=3--2 and 2--1 survey of the same 21 sources
taken in beam-switched mode shows dense compact circumstellar material associated with
only 2 sources, HT Lup and Sz 73.

\item{--} Only a single source, IM Lup (Sz 82), has been found with
clear evidence for a gas-rich disk. Modeling of the line intensity and
line profile indicates a large disk with radius $\sim$600 AU and
inclination 20$^\circ$. This source will be a prime target for future
interferometer observations in the southern sky, especially with the
{\it Atacama Large Millimeter Array}.
\end{description}
The lack of detection of gas disks in Lupus in the current survey does
not imply that the gas has been dissipated from these disks; rather,
it shows the limitations of single-dish CO line observations to reveal
them.  Future searches need to either focus on
high density single-dish tracers such as HCO$^+$ $J$=4--3 or HCN
$J$=4--3 to distinguish the warm, dense gas in the disk from the cold,
diffuse cloud material or use newly commisioned interferometers such as the SubMillimeter Array (SMA).  From comparison with other disks in Taurus
\citep{Thi04,Greaves04}, these lines are expected to have peak
intensities of at most 0.1 K in a single-dish 15$''$ beam and will be very
difficult to detect.  Increased sensitivity to compact emission is
only possible with interferometer studies.  Interferometric surveys
will therefore be essential to determine the fraction of gas-rich
disks in Lupus and thus the timescale for gas dissipation in this star
forming region compared with other regions such as Taurus.
%\end{description}

%___________________________________________________________________

\begin{acknowledgements}

The authors are grateful to Remo Tilanus for carrying out part of the
observations and to Karl Stapelfeldt, Debbie Padgett, Glenn Schneider and Jean-Charles
Augereau for communicating information about the HST images of IM Lup
before publication.  TvK and astrochemistry at Leiden Observatory are
supported by a Spinoza grant and by grant 614.041.004 from the Netherlands Organization for Scientific Research (NWO).

\end{acknowledgements}
\bibliographystyle{../bibtex/aa}
\bibliography{../biblio}

%\Online
%  \begin{figure*}
%  \centering
%       \includegraphics[width=140pt]{./online/12co/hmlup.ps}
%       \includegraphics[width=140pt]{./online/12co/holup.ps}
%      \includegraphics[width=140pt]{./online/12co/htlup.ps}
% \includegraphics[width=140pt]{./online/12co/sz73.ps}
%       \includegraphics[width=140pt]{./online/12co/sz84_2.ps}
%\includegraphics[width=140pt]{./online/12co/sz96.ps}

%     \caption{$^{12}$CO  3--2,}
%              \label{gqlup}
%   \end{figure*}
%  \begin{figure*}
%  \centering
%       \includegraphics[width=140pt]{./online/12co/sz107.ps}
%       \includegraphics[width=140pt]{./online/12co/sz109.ps}
%       \includegraphics[width=140pt]{./online/12co/sz110.ps}
% \includegraphics[width=140pt]{./online/12co/sz117.ps}
%\includegraphics[width=140pt]{./online/12co/sz123.ps}
%      \includegraphics[width=140pt]{./online/12co/v908sco.ps}
%      \caption{$^{12}$CO  3--2 lines of selected sources in Lupus, 
%with off-positions 30$''$ south and east of source.}
%              \label{sz107}
%   \end{figure*}

\end{document}